# Superconducting energy gap of *2H*-NbSe₂ in phonon spectroscopy

*F. Weber[1], S. Rosenkranz[2], R. Heid[1], A. H. Said[3]*

[1]Karlsruhe Institute of Technology, Institute for Solid State Physics, 76021 Karlsruhe, Germany
[2]Materials Science Division, Argonne National Laboratory, Argonne, Illinois, 60439, USA
[3]Advanced Photon Source, Argonne National Laboratory, Argonne, Illinois, 60439, USA

**We present a high energy-resolution inelastic x-ray scattering data investigation of the charge-density-wave (CDW) soft phonon mode upon entering the superconducting state in *2H*-NbSe₂. Measurements were done close to the CDW ordering wavevector $q_{CDW}$ at $q = q_{CDW} + (0,0,l)$, $0.15 \leq l \leq 0.5$, for $T = 10$ K (CDW order) and $3.8$ K (CDW order + superconductivity). We observe changes of the phonon lineshape that are characteristic for systems with strong electron-phonon coupling in the presence of a superconducting energy gap $2\Delta_c$ and from which we can demonstrate an l-dependence of the superconducting gap. Reversely, our data imply that the CDW energy gap is strongly localized along the c* direction. The confinement of the CDW gap to a very small momentum region explains the rather low competition and easy coexistence of CDW order and superconductivity in *2H*-NbSe₂. However, the energy gained by opening $\Delta_{CDW}$ seems to be too small to be the driving force of the phase transition at $T_{CDW} = 33$ K, which is better described as an electron-phonon coupling driven structural phase transition.**

The interplay between superconductivity and other ground states of solids is one of the most challenging topics in condensed matter physics documented, e.g., in the intense research on high-temperature superconductivity [1]. It is now evident that a number of layered materials exhibiting charge-density-wave (CDW) order, a periodic modulation of the charge carrier density and the atomic lattice, become superconducting once CDW order is suppressed.[2-6] Classic examples are the members of the transition-metal dichalcogenide family MX₂, where M = Nb, Ti, Ta, Mo and X = S, Se, which show a large diversity of CDW ordered phases competing with SC [6,7]. One expects that the maximum $T_c$ occurs close to the critical value of the extrinsic parameter for which $T_{CDW}$ is suppressed to zero if the electron-phonon-coupling (EPC) of the CDW soft phonon mode is the main mediator of superconductivity.

*2H*-NbSe₂ is special in that it supports CDW order ($T_{CDW} = 33$ K [8,9]) and superconductivity ($T_c = 7.2$ K [10]) simultaneously at ambient pressure. For decades, the driving mechanism of the CDW transition was hotly debated since the Fermi surface nesting required for a 2D analogue of the Peierls transition was not observed. Various alternative weak-coupling scenarios were proposed [11,12]. However, the strong coupling model with wave vector dependent electron-phonon coupling (EPC), originally introduced in the late 1970s [13,14], was recently proven to explain many of the puzzling experimental results [15-19]. *2H*-NbSe₂ can therefore be regarded as a prototypical strong-coupling model system for materials with charge order in more than one dimension, which are governed by both the properties of the EPC matrix elements and the electronic band structure.

While CDW order in *2H*-NbSe₂ can be easily suppressed by intercalation [18] or applied pressure [20,21], superconductivity is less sensitive with respect to these extrinsic parameters. Angle-resolved photoemission spectroscopy (ARPES) showed that the CDW energy gap $\Delta_{CDW}$ opens only on a small part of the Fermi surface within the $(h,k)$-plane of *2H*-NbSe₂ [22]. However, nearly 20% of the total electron-phonon-coupling (EPC) in *2H*-NbSe₂ is located in the CDW soft phonon mode [19] and one does not observe a corresponding increase of the superconducting transition temperature when the CDW gap is suppressed by intercalation [23]. Under pressure, $T_c$ increases but the maximum value is not observed for the critical pressure $p_c \approx 5$ GPa, for which CDW order is suppressed to zero temperature, but near a pressure of 10 GPa [24]. Hence, the interplay between the two phases in *2H*-NbSe₂ and, in particular, the relevance of the EPC of the CDW soft phonon mode for superconductivity remain puzzling.

Here, we report an investigation of the superconductivity-induced changes of the phonon lineshape of the low energy phonon branch dispersing out of the CDW ordering wavevector $q_{CDW}$ on cooling from $T = 10$ K to $3.8$ K. We find signatures of the opening of the superconducting gap $2\Delta_c$ and extract a wavevector dependent value $2\Delta_c(q)$ using a model published by Allen *et al.* [25], which we successfully used in previous investigations of other superconductors [26,27]. Our results demonstrate that these phonons with wavevectors $q = q_{CDW} + (0,0,l)$ couple to ungapped parts of the Fermi surface and, hence, the CDW gap is strongly localized also along $(0,0,l)$.

The phonon measurements were carried out using the HERIX spectrometer at sector 30 at the Advanced Photon Source (APS) at Argonne National Laboratory. The beam on the sample was focused to $35x15$ μm². The incident energy was 23.724 keV [28] and the horizontally scattered beam was analysed by a set of spherically curved silicon analysers (Reflection 12 12 12) [29]. The full width at half maximum (FWHM) of the energy and momentum resolution was about 1.5 meV and 0.066 Å⁻¹, respectively. The components ($Q_h$, $Q_k$, $Q_l$) of the scattering vector are expressed in reciprocal lattice units (r.l.u.) ($Q_h$, $Q_k$, $Q_l$ = (h*2π/a, k*2π/a, l*2π/c) with the lattice constants a = b = 3.443 Å and c = 12.55 Å of the hexagonal unit cell, space group P6₃/mmc. Energy scans at constant momentum transfer **Q** = **τ** + **q**, where **τ** is a reciprocal lattice point and **q** the reduced wave vector, were performed in the Brillouin zone adjacent to **τ** = (3, 0, 1), i.e. **Q** = $(3 − h, 0, 1)$. We used a high-quality single crystal sample of about 50 mg ($2 \times 2 \times 0.05$ mm³) already used in previous investigations [15,16] with $T_{CDW} = 33$ K [30] and a superconducting transition temperature of $T_C = 7.2$ K [9]. The sample was mounted in a closed cycle refrigerator with a minimum temperature of $T = 3.8$ K. Good counting statistics were achieved by measuring up to 28 minutes per energy point.

Raw IXS data at a momentum transfer $\boldsymbol{Q} = (2.671,0,l)$, $0.15 \leq l \leq 0.5$, taken above and below $T_c$ are shown in Figure 1a. This momentum transfer corresponds to the CDW ordering wavevector $\boldsymbol{q}_{CDW}$ in the reduced Brillouin zone, where the soft phonon mode of the CDW transition at $T_{CDW} = 33$ K is observed for $l = 0$ [15]. We generally find three components in the data: (1) a resolution limited elastic line, (2) a broad phonon dispersing at 4 meV $< E <$ 6 meV and (3) another phonon mode near 10 meV. The elastic line is largest for $l = 0.15$ because we already catch the tail of the CDW superlattice peak centered at $l = 0$ within the finite momentum resolution of the spectrometer. Phonon measurements at even smaller values of $l$ are, therefore, not possible anymore. The mode near 10 meV is an optic phonon, which is not strongly involved in the CDW transition [15,16]. The broad low energy phonon extrapolates to the CDW soft phonon mode for $l = 0$ and we will, for simplicity, refer to the modes at finite $l$ as the soft mode as well. In fact, density-functional-perturbation theory (DFPT) predicts a strong electron-phonon coupling for the complete soft phonon branch dispersing at $\boldsymbol{q} = \boldsymbol{q}_{CDW} + (0,0,l)$, $0 \leq l \leq 0.5$ (Suppl. Fig. 1).

We plot the observed soft mode dispersion at $T = 10$ K for $\boldsymbol{q} = \boldsymbol{q}_{CDW} + (0,0,l)$ in Figure 1b. We find only weakly wavevector dependent energies, a signature of the quasi two-dimensional structure with weak bonding along the $c$ axis. Comparing the soft mode dispersion to previous measurements at $T_{CDW} = 33$ K [16] (solid line in Fig. 1b), we find hardened energies at $T = 10$ K and $l = 0.15$. The changes on cooling from $T_{CDW}$ down to $T = 10$ K decrease for increasing $l$ and are negligible at $l = 0.5$.

Here, we want to focus on the properties of the soft phonon mode at different $l$. Superconductivity-induced changes are most obvious for $l = 0.3$ and 0.4 (Fig. 1a): Data taken in the superconducting phase ($T = 3.8$ K) show increased scattering intensities for 2 meV $\leq E \leq$ 3 meV ($l = 0.3$) and 2 meV $\leq E \leq 4$ meV ($l = 0.4$), whereas we observe decreased intensities at energies $E \approx 5 - 6$ meV. As we will explain below in detail such a redistribution of spectral weight is typical for phonons that interact strongly with electrons on the Fermi surface upon entering the superconducting phase as has been reported previously in other systems [26,27]. Using the same analytical tools based on the theoretical work of Allen et al. [25] as used in the previous studies, we can extract values of the superconducting energy gap and obtain new insights on the interplay between the CDW soft-mode and superconductivity in $2H$-NbSe$_2$.

In Figure 2, we demonstrate the analysis of the normal state data taken at $\boldsymbol{Q} = (2.671,0,1.3)$ and how we model the superconductivity-induced changes. First, we show a standard analysis of IXS raw data at $T = 10$ K (Fig. 2a), which were fitted using a resolution limited pseudo-Voigt function for the elastic line and damped harmonic oscillator (DHO) functions [31], convoluted with the experimental resolution, $\Delta E = 1.5$ meV (full width at half maximum, horizontal bar in Fig. 2a), for each phonon peak. The energy $\omega_q$ of the damped phonon is obtained from the fit parameters of the DHO function by $\omega_q = \sqrt{\widetilde{\omega}_q^2 - \Gamma^2}$ [32], where $\Gamma$ is the imaginary part of the phonon self-energy and $\widetilde{\omega}_q$ the phonon energy renormalized only by the real part of the phonon self-energy.

According to the theory developed by Allen et al. [25] for a phonon profile at $\boldsymbol{q} \neq 0$ in a superconductor, the part of the phonon low energy tail with $E < 2\Delta_c$ is pushed up in energy to form a narrow spike at $2\Delta_c$. Hence, we need only consider the soft phonon mode because only this mode has a finite intensity at $E \leq 2\Delta_c \leq 2.5$ meV [22]. Calculations based on the framework of Ref. [25] for the soft mode at $\boldsymbol{Q} = (2.671,0,1.3)$ are shown in Fig. 2b. The experimental input parameters for the theory are $\widetilde{\omega}_q$ and the damping ratio $\Gamma/\widetilde{\omega}_q$. If we set $2\Delta_c = 0$ we replicate the fitted DHO function shown in Fig. 2a (dash-dotted line in Fig. 2b). Setting $2\Delta_c$ to a finite value creates a narrow spike in the phonon lineshape at $E = 2\Delta_c$ (dashed red line in Fig. 2b), which is washed out by the finite energy resolution in our IXS experiment (solid red line in Fig. 2a). This theory predicts that the opening of the superconducting gap will be visible as an increase (decrease) of the scattering intensity on the low (high) energy side of the phonon peak, which corresponds very well to the observed changes on entering the superconducting phase (Fig. 1a). As we discussed previously [33], the difference between the phonon lineshape in the normal state ($\Delta_c = 0$) and that in the superconducting phase ($\Delta_c > 0$) shows a characteristic sequence of positive – negative – positive values with increasing energies starting from zero (Fig. 2c).

The corresponding experimental differences between the IXS data taken at $T = 10$ K and 3.8 K are shown in Fig. 3a along with curves calculated as outlined above. Calculations have been done for various values of $2\Delta_c$. Difference lines yielding the best description of the experimental data are shown as solid lines in Fig. 3a. We find the best agreement between theory and experiment for $2\Delta_c = 1.2$ meV at $l = 0.15$, $2\Delta_c = 1.8$ meV at $l = 0.3$ and $2\Delta_c = 2.4$ meV for $l = 0.4$ and 0.5 (Fig. 3b). Although this is not a high-precision determination, we can demonstrate the significance of the increase of $2\Delta_c$ by comparison: For instance, if we take the input parameter for $l = 0.15$ but set $2\Delta_c = 2.4$ meV (green dashed line in Fig. 3a), the predicted effect is clearly too large compared to experiment. Similarly, taking $2\Delta_c = 1.2$ meV for $l = 0.4$ and 0.5 (blue dashed lines in Fig. 3a) underestimates the observed effects.

Our results show that the Fermi surface parts connected by the wavevectors of the soft phonon with finite $l$ are not gapped by the onset of CDW order and, hence, the strong EPC of these modes is available for phonon mediated superconductivity even in the CDW ordered phase. This explains the negligible effect of suppressing CDW order on the superconducting transition temperature because the EPC bound in the soft mode directly at $\boldsymbol{q}_{CDW}$ is very small, i.e. only a few per cent.

The electronic contribution to the phonon linewidth $\Gamma$ integrates over all scattering contributions from electronic states at the Fermi level connected by the respective phonon wavevector $q$ [34]. Hence, we need to consider the Fermi surface topology of *2H*-NbSe$_2$ for a more detailed analysis of our results: Several bands are crossing the Fermi energy [17,22,35] but we showed previously that the EPC of the soft mode, i.e. its linewidth Γ, originates only from electronic scattering between Nb-derived states around the *K* point [16]. These states form two cylindrical Fermi surfaces with only little dispersion along the out-of-plane direction $k_z$ [15,36] and the CDW gap $\Delta_{CDW}$ was found to appear only on several points, well separated from each other, on these Fermi surface cylinders for $k_z = 0$ [22]. These CDW hot spots are connected by $q_{CDW}$. If our phonons would only probe electronic states with $k_z = 0$ there could not be any further superconductivity induced effects because of the large $\Delta_{CDW} \geq 4.5$ meV [22]. In reality, the soft mode probes electronic states at arbitrary $k_z$ which may or may not be gapped by $\Delta_{CDW}$ and, hence, may not or may develop a finite $\Delta_c$. Using IXS we cannot investigate the soft mode at $l = 0$ for $T < T_{CDW} = 33$ K [15]. Our measurements at $0.15 \leq l \leq 0.5$, however, effectively probe the same electronic states but connected by $\Delta k_z = l$. Our experimental results (Figs. 1-3) clearly show the opening of $2\Delta_c$.[1] and demonstrate that the probed part of the Fermi surface features a superconducting gap. DFPT results show that Γ of the soft mode at $q = q_{CDW} + (0,0,l)$ with $0 < l \leq 0.5$ originates from the same electronic scattering paths as the linewidth of the soft mode for $l = 0$ (Suppl. Fig. 1d). Therefore, we conclude that $\Delta_{sc}$ and $\Delta_{CDW}$ develop on the same in-plane positions of the Fermi surface but at different $k_z$ and that $\Delta_{CDW}$ is localized in three dimensions. Our results explain why superconductivity and CDW order only weakly compete and coexist because the small phase space, in which $\Delta_{CDW}$ appears, does not significantly reduce the overall electron-phonon coupling constant that is driving phonon-mediated superconductivity. Furthermore, the observations made here that only a small electronic phase space is gapped by the onset of CDW order agrees with recent conclusions of a strong EPC driven phase transition based on observations of strong EPC [15,16], absence of Fermi surface nesting [17,36,37] and an electron-hole asymmetric CDW gap [18,38].

Recently, thin sheets and single layers of 2H-NbSe$_2$ have been studied as model systems for the effect of dimensionality on correlated electronic phases [39,40]. In a 2D material there is no $k_z$ dependence and, hence, we would expect that the complete Fermi surface connected by the in-plane CDW ordering wavevector is gapped for $T < T_{CDW}$. $T_c$ should decrease in 2D-NbSe$_2$ because the coupling of the soft mode is not available for phonon-mediated superconductivity anymore. Reversely, CDW order should be stabilized by the enhanced Fermi surface nesting in 2D. This reasoning is in excellent agreement with the reported temperature-dimensionality phase diagram [40] and our results provide a microscopic understanding of the behavior in single layer NbSe$_2$.

In conclusion we have reported superconductivity-induced changes of the lineshape of phonons closely related to the CDW soft phonon mode in *2H*-NbSe$_2$, from which we can demonstrate a $k_z$-dependence of the superconducting gap $\Delta_c$. Our experiments demonstrate that the strong EPC of these low energy modes is available for phonon-mediated superconductivity even in the CDW ordered phase and, hence, suppression of CDW order has only little effect on $T_{sc}$ itself. Our observations imply that the CDW energy gap $\Delta_{CDW}$ is strongly localized along $k_z$, a result which could not be observed by two-dimensional probes such as angle-resolved photoemission spectroscopy [22]. Our results impose fundamental questions about the interplay of EPC and anharmonicity, which need further theoretical investigations for a deeper understanding of the phase competition/coexistence in transition-metal dichalcogenides as well as other superconducting materials with competing ground states.

F.W. was supported by the Helmholtz Society under contract VH-NG-840. Work at Argonne was supported by U.S. Department of Energy, Office of Science, Office of Basic Energy Sciences, under Contract No. DE-AC02-06CH11357.

---

[1] The values of $2\Delta_c$ plotted in Figure 3b are, in general, the sum of two gaps $\Delta_1$ and $\Delta_2$ developing at the two points in momentum space connected by the phonon wavevector $q$ and averaged over all parts of the Fermi surface connected by $q$.

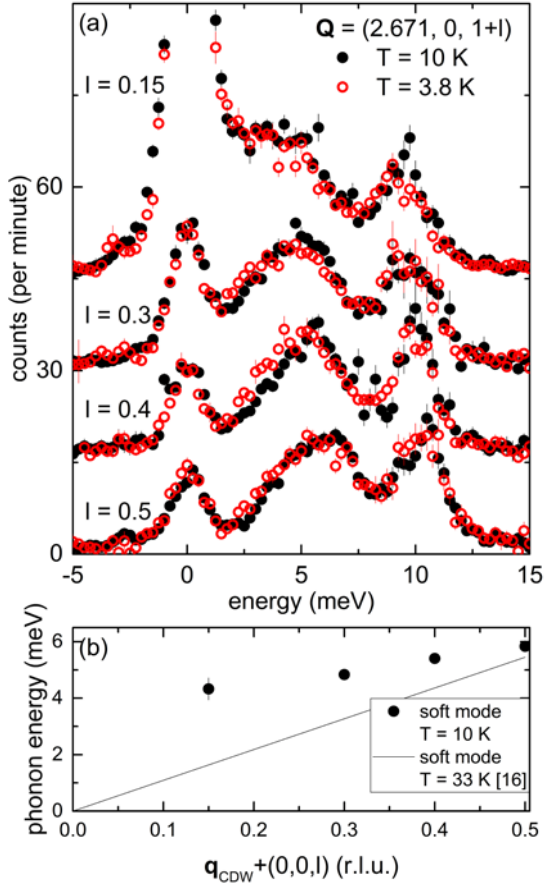
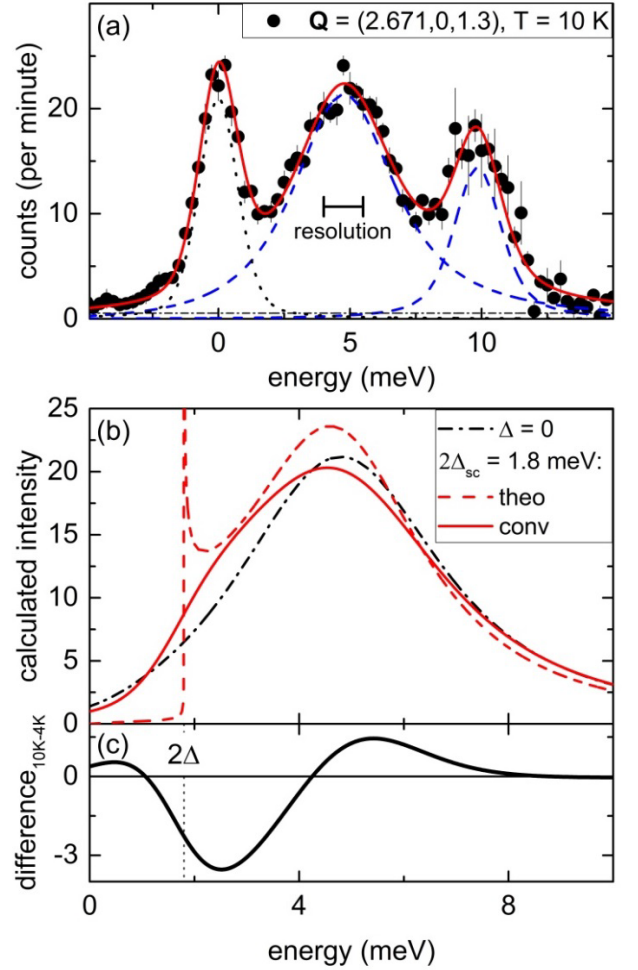

**Figure 1:** (color online) *(a)* Raw data of inelastic x-ray scattering in *2H*-NbSe2 at $T = 10$ K (dots) and 3.8 K (circles), i.e. above and below the superconducting transition temperature $T_c = 7.2$ K. Data were taken at $Q = (2.671, 0, l)$ and $0.15 \leq l \leq 0.5$. *(b)* Observed phonon energies at $T = 10$ K (dots) in comparison to the reported dispersion of the soft mode at $T = T_{CDW} = 33$ K (line) [16].

**Figure 2:** (color online) *(a)* Exemplarily analysis of IXS data (dots) taken in the non-superconducting phase of *2H*-NbSe2. The solid line is a fit using damped harmonic oscillator functions for the phonons (dashed blue lines), a resolution limited pseudo-voigt function for the elastic line (dotted line) and a constant background (dashed-dotted line). The vertical bar represents the experimental resolution of 1.5 meV (FWHM). *(b)* Calculation of the spectral weight distribution in the superconducting phase with an energy gap of $\Delta = 0.9$ meV. The solid black line corresponds to the DHO fit of the lower energy phonon in panel (a). The dashed red line is the calculation based on the theory of Allen et al. [Allen1997] and the solid red line includes the effect of the finite experimental resolution. *(c)* Difference between the non-superconducting and calculated superconducting phonon lineshape shown in panel (b). The position of the superconducting energy gap $2\Delta = 1.8$ meV is indicated by the vertical dashed line.

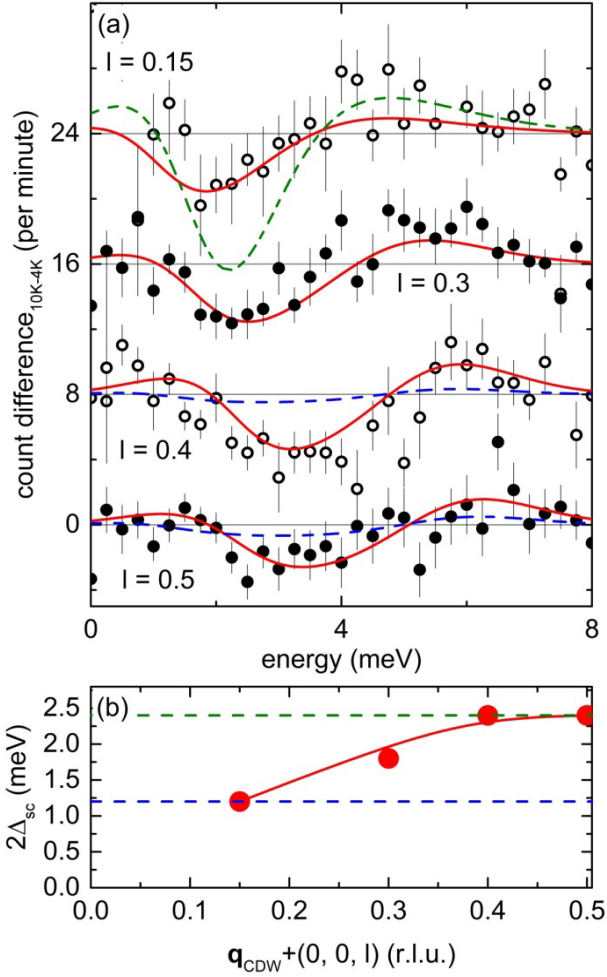

**Figure 3:** (color online) *(a)* Difference curves of the observed intensities at $Q = (2.671, 0, l)$, $0.15 \leq l \leq 0.5$, at $T = 10$ K and 3.8 K. Data for $l < 0.5$ are offset for clarity (offsets indicated by the thin horizontal lines). Red solid lines are calculated differences (see text and Figure 2) employing $\Delta = 0.6$ meV ($l = 0.15$), 0.9 meV ($l = 0.3$) and 1.2 meV ($l = 0.4$ and 0.5). Dashed lines are results calculated with $\Delta = 1.2$ meV ($l = 0.15$) and 0.6 meV ($l = 0.4$ and 0.5). *(b)* $\Delta$ values as function of $l$ (dots) yielding the best agreement between experiment and theory. The solid line is a guide to the eye assuming $\Delta = 0$ at the CDW superlattice wavevector, i.e. at $l = 0$. The dashed horizontal lines indicate the largest and smallest $\Delta$ values used, for which calculated differences are also plotted in panel *(a)* for comparison.